\documentclass[a4paper,12pt]{article}

\def\ad   {a^{\dagger}}

\def\HP   {\hat {P}}

\def\al   {\alpha}

\def\bet  {\beta}

\def\HH {\hat H}

\def\HP {\hat P}

\def\HH {\hat H}

\def\oc {\overline c}

{\it}
{\it}
{\it}
{\it}
{\it}
\def\exp {\rm {e}}{\it}

\def\bet {\beta}

\def\eps {\epsilon}

\def\om {\omega}
\def\Om {\Omega}

\usepackage{times}
\usepackage{graphicx}
\begin{document}
\title{ A new single-particle basis for nuclear many-body calculations.}
\author{G. Puddu\\
       E-mail: giovanni.puddu@mi.infn.it\\
       Dipartimento di Fisica dell'Universita' di Milano,\\
       Via Celoria 16, I-20133 Milano, Italy}
\maketitle
\begin {abstract}
      Predominantly, harmonic oscillator single-particle wave functions
      are the choice as a basis in ab-initio nuclear many-body calculations.
      These wave-functions,
      although very convenient in order to evaluate the matrix elements of
      the interaction  in the laboratory frame, have a too fast fall-off at large distances.
      In the past, in alternative to the harmonic oscillator, other single-particle
      wave functions have been proposed. In this work we propose a new single-particle
      basis, directly linked to the nucleon-nucleon interaction. This new basis is orthonormal
      and complete, has the proper asymptotic behavior at large distances and does
       not contain the continuum
      which would pose severe convergence problems in nuclear many body calculations.
      We consider the newly proposed NNLO-opt nucleon-nucleon interaction,
      without any renormalization.
      We show that unlike other basis,  this single-particle representation has a
      computational cost similar to the harmonic oscillator basis with the same  
      space truncation and it gives lower energies for ${}^6He$ and ${}^6Li$.
\par\noindent
{\bf{Pacs numbers}}: 21.10.-k,21-60.Cs, 24.10.Cn
\vfill
\eject
\end{abstract}
\section{ Introduction.}
\bigskip
\par
      New theoretical methods and advanced computational facilities, have made possible in recent years
      to tackle the most fundamental problem in nuclear many-body theories. That is, the description of 
      nuclear properties starting from the nucleon-nucleon interaction. 
      Several modern nucleon-nucleon interaction and NNN interactions,
      based on chiral  perturbation theory are nowadays available (refs. [1]-[5]).
      These interactions are the input to modern many-body methods in order to extract nuclear observables.
      To mention a few, the no-core shell model (refs. [6],[7] and ref. [8] for a recent review),
      the coupled-cluster approach (refs. [9],[10] and for a review ref.[11]), the similarity renormalization
      group method (cf. ref.[12] for a recent review) and the Self-Consistent Green-s Function method
      (ref.[13].
      By large all these methods use the harmonic oscillator (h.o.) wave functions
      as the single-particle basis. Effects due to truncation of the Hilbert space are addressed using
      robust infrared extrapolation techniques (ref. [14]).
      Only recently there has been a systematic attempt to consider an alternative basis, namely the
      Coulomb-Sturm basis
      (refs. [15],[16])  which
      has the following properties. It is orthonormal and complete, it does not have continuum states and it has the 
      proper asymptotic behavior in coordinate space at large distances (i.e. it falls off as $\exp^{-\mu r}$).
      In the past we have considered 
      an alternative basis very similar to the one we propose in this work,
      which however has been used for a very simple model and does not have the proper
      asymptotic behavior (ref.[17]) since it has a Gaussian falloff.
      One of the reasons the Coulomb-Sturm basis has been used, was that
      quantities like root mean square radii, quadrupole moments and transition rates are sensitive to
      large distances.
      The Coulomb-Sturm basis did improve in the description of these quantities.
      Moreover the use of a basis with the correct asymptotic behavior is highly desirable in the description 
      of halo nuclei (ref.[16]).
      Moreover it has a computational cost similar to the harmonic oscillator with the same space truncation.
      However, the Coulomb-Sturm
      basis has an important shortcoming as pointed out in ref. [16].
      Namely, the energies produced in many-body calculations are much higher
      than the corresponding ones obtained with  the harmonic oscillator basis.
      More recently, the natural orbit basis has been considered as a candidate for the description of halo 
      systems (ref.[18]). Its main advantage is that it produces energies lower than the ones
      obtained with the h.o. basis. However this basis requires a preliminary shell model calculation. While
      for light systems this may pose no problem, it could be computationally demanding for heavier nuclei.
\par
      Also, in the past some no-core shell-model calculations have been performed using a Wood-Saxon basis.
      However the parameters of the Wood-Saxon potential have to be varied in order to minimize the
      shell model results for the ground-state energy (ref.[19]).
\par
      It is the purpose of this work
      to propose a new basis which seems to be free of the shortcoming of the Coulomb-Sturmian basis. 
      If we desire a better basis
      than the harmonic oscillator, it should, with a comparable numerical effort, lead to  lower energies in many-body
      calculations. The basis we propose is essentially the basis of ref. [17] properly corrected in order 
      to have the appropriate asymptotic behavior in coordinate space. Contrary to the Coulomb-Sturm basis
      it has its roots in the NN interaction. 
      In what follows, it should be kept in mind that our primary purpose is the description of nuclei where
      the long range part of the wave function is important, although it could have an impact for nuclei
      in the valley of stability.
      The basic reasoning behind our basis is as follows.
      Consider the Hamiltonian in the center of mass system for  $A$ particle interacting with a potential $V_{ij}$, 
      $H=\sum_{i<j}H_{ij}=\sum_{i<j}((\vec p_i-\vec p_j)^2/2mA +V_{ij})$ 
      and let us diagonalize $H_{ij}$. In this work we take $A=2$ in order to obtain the single-particle basis
      to be used in many-body calculations.
      This may not be the best choice for ${}^4He$ since it is  a very compact object. 
      Let us consider the ground-state of  Deuterium,
      let us discard the $D$-part of the wave-function and consider only 
      the $S$ part. This wave function depends on the relative momentum of the neutron and proton 
      and it is not localized in coordinate space. To achieve  localization,  we consider the
      full wave function which contains also  the wave-function (in an $S$ state) of the 
      center of mass of the system. The center of mass part can be used to localize
      the system. Arguing for simplicity in coordinate space (although we work in momentum space), the total wave 
      functions depends on $|\vec r_n-\vec r_p|$ and $|\vec r_n+\vec r_p|$ hence on $r_n,r_p$ and
       $\cos( \theta_{np})$
      the angle between the position vectors of the neutron and the proton. We can analyze the  $\cos( \theta_{np})$
      dependence in therms of Legendre polynomials and relate these to the spherical harmonics  of the angular
      coordinates of the neutron and the proton. The net result is that the Deuterium wave function is rewritten
      as a linear combination of products of functions  $F_l(r_n,r_p)$, which we will discuss in detail later
      and the spherical harmonics of $\theta_{n}$ and $\theta_{p}$.
      We can diagonalize these functions $F_l(r_n,r_p)$ on a lattice and obtain the total Deuterium wave function
      in terms of single-particle
      wave functions of the neutron and the proton. These single-particle  wave functions are orthonormal, they are 
      complete by  construction, they  have the proper
      asymptotic behavior at large distances (for a convenient choice of the center of mass wave function)
      and do not contain continuum states. Moreover they have a very useful additional 
      feature. By controlling the space extent of the center of mass we can "squeeze" or "spread" in space 
      the single-particle wave functions. This basis can be used in many-body calculations, although it
      has been constructed
      from the S-part of the Deuterium ground-state. Note that in principle we could construct a whole
      family of basis by weighting
      properly the kinetic energy term in $H_{ij}$.
      We call this new basis Localized Deuteron Basis (LDB).
\par
     In the cases discussed in this paper this set of single-particle
      wave functions produces  energies of better or the same quality  obtained using an optimized h.o. basis, except
      the case of ${}^4He$.   
      We have not carried out the optimization suggested in ref. [15], namely the optimization
      of the radial wave functions
      for each single-particle angular momentum. The optimization of our set is performed only modifying 
      the "tail"  of all radial wave functions. We expect that the implementation of the optimization for
      every single-particle angular momentum will improve the energies even more.
      The use of radial wave functions other than the harmonic oscillator poses the additional problem
      of the evaluation of the matrix elements in the laboratory frame of the two-body interaction. This problem
      is solved in the h.o. basis by the Talmi-Moshinky brackets (ref.[20]). For basis other than the h.o.
      wave functions,
      the problem can be addressed using the vector brackets, discussed in refs. [21]-[24]. Here we use the expansion
      of our basis in the harmonic oscillator basis as done in ref. [15], using a rather large number
      of major oscillator shells.
      The many-body approach we use is the Hybrid-Multideterminant method (HMD) (refs.[25],[27]), whereby the nuclear
      wave function is expanded as a linear combination of a rather large number of Slater determinants.
      This paper is organized as follows. In section 2 we describe in detail the construction of the basis
      and some of its properties. In section 3 we compare the harmonic oscillator basis with the
      one we propose with a brief recap of the many-body method that we use in subsection (3.a). 
      In section 4 we present some conclusions.
\vfill
\section{ Choice of the single-particle basis}. 
\bigskip
\par
       Let us start by constructing in momentum space the ground-state wave function of
       Deuterium. As well known it has an $L=0$ component and an $L=2$ part. Let us isolate
       the $L=0$ part and let us call the radial part $u(k)$, where $\vec k=(\vec k_1-\vec k_2)/2$,
       $ \vec k_1$ and $\vec k_2$ being the momenta of the nucleons. Let us discard completely the
       $L=2$ part of the deuterium wave function and let us construct the following wave-function
$$
\Psi(\vec k_1,\vec k_2)= u(k) \Theta(K)
\eqno(2.1)
$$
       where  $\Theta(K)$ is for the time being an unspecified scalar wave function of the total
       momentum $\vec K=\vec k_1+\vec k_2$. The wave function in eq.(2.1) is normalized to $1$.
       In coordinate space, the role of $\Theta$ (or better
       of its Bessel-Fourier transform) is to localize the Deuterium. 
       The right-hand side of eq.(2.1) depends on the relative orientation of $\vec k_1,\vec k_2$
       only through the cosine of the relative angle $\theta_{12}$ between the momenta
       $\vec k_1,\vec k_2$.
       We can analyze the r.h.s of eq.(2.1) using Legendre polynomials $P_l(\cos(\theta_{12}))$
       and write
$$
\Psi(\vec k_1,\vec k_2)= \sum_{l=0}^{\infty} f_l(k_1,k_2) P_l(\cos(\theta_{12}))
\eqno(2.2)
$$
       where
$$
f_l(k_1,k_2)= (l+{1\over 2})\int_0^{\pi} d\theta_{12} \sin(\theta_{12})  P_l(\cos(\theta_{12}))
\Psi(\vec k_1,\vec k_2)
\eqno(2.3)
$$
       Using the familiar addition theorem of the spherical harmonics $Y_{lm}$ we obtain       
$$
\Psi(\vec k_1,\vec k_2)= \sum_{lm} f_l(k_1,k_2){4\pi \over 2 l+1} Y^*_{lm}(\hat k_1)Y_{lm}(\hat k_2)
\eqno(2.4)
$$
       Using the normalization condition on the wave function $\Psi(\vec k_1,\vec k_2)$ we obtain
$$
\int k_1^2 k_2^2 dk_1 dk_2 \sum_{lm} f_l(k_1,k_2)^2 \big ({4\pi\over 2l+1}\big )^2=1
\eqno(2.5)
$$
       Let us now discretize the lab. coordinates $k_1,k_2$ on a mesh of spacing $\Delta k$ and let us 
       define the eigenvalue problem for the real symmetric matrix $k_1 f_l(k_1,k_2) k_2\equiv M_{i,j}$
       where $i,j$ refer to the position on the lattice of $k_1$ and $k_2$
$$
k_1 f_l(k_1,k_2) k_2\equiv M_{i,j}=\sum_{n=0} v^{(l)}(i,n)\eps^{(l)}_n v^{(l)}(j,n)
\eqno(2.6)
$$
      For later convenience, the index $n$ which labels the eigenvalues takes the values $0,1,2,..$ and has the role 
      of radial quantum number.
      We reorder the eigenvalues $\eps^{(l)}_n,\; n=0,1,..$ for a fixed $l$ so that $|\eps^{(l)}_n|$ decrease
      with increasing $n=0,1,2,..$.
      We obtain  the following expansion 
$$
\Psi(\vec k_1,\vec k_2)=\sum_{l,n} {v^{(l)}(i,n)\over k_1}\eps^l_n {v^{(l)}(j,n)\over k_2} {4\pi\over 2l+1}
\sum_m Y^*_{lm}(\hat k_1)Y_{lm}(\hat k_2)
\eqno(2.7)
$$
      Therefore the $L=0$ part of the Deuterium wave-function  has been recast as an expansion 
      of single-particle wave-functions in the lab. frame. Defining
$$
\phi_{n,l,m}(k)= {v_l(i,n)\over k\sqrt{\Delta k}} Y_{lm}(\hat k)\equiv{ Q_{n,l}(k)\over k}Y_{lm}(\hat k)
\eqno(2.8)
$$    we obtain a set of radial single-particle wave functions in the lab. frame ${ Q_{n,l}(k)\over k}$.
      These wave functions can be used as a single-particle basis to perform many-body calculations,
      much in the same way of the Coulomb-Sturm  wave functions. The wave functions defined by eq.(2.8) are
      however more natural in nuclear many-body calculations.
\par 
      There are several points to be analyzed. First of all, we have obtained a discrete and complete set
      of single-particle wave functions in the lab. frame, directly linked to the underlining NN-interaction.
      Completeness stems from the unitarity of the eigenvectors $v$. Also, since we can localize nucleons
      with the appropriate $ \Theta(K)$, this set does not contain the continuum, which would pose severe problems
      of  convergence in many-body calculations.
      Moreover the normalization of the total wave-function $\Psi$ gives
$$
\sum_{l,m,n} \big [{4 \pi \Delta k \eps^{(l)}_n\over 2 l+1}\big ]^2=1
\eqno(2.9)   
$$
      Hence the quantities
$$
p_{n,l}= [{4 \pi \Delta k \eps^{(l)}_n\over 2 l+1}]^2
\eqno(2.10)
$$
       give the 'probability' of a nucleon  being
      in the single-particle state characterized by the quantum numbers $n,l,m$. Since the series has to converge,
      we expect the $p_{n,l}$ to decrease for large $n,l$. Hopefully the most important part of the deuterium
      wave function is expanded as a sum (in a shell model fashion) of "few" single-particle wave functions
      in the lab. frame. To fix the ideas, let us consider the recent NN interaction NNLO-opt recently
      introduced in ref. [4]. Let us extract and normalize the $S$ part of the ground-state wave-function
      (in this work no renormalization step is taken on the NN interaction). The function
      $\Theta(K)$ is taken to be the Bessel-Fourier transform of a localizing center of mass wave-function.
      For the nuclear case in order to have an asymptotic behavior of the type $\exp^{-\mu r}$ at large distances
      we considered the Bessel-Fourier transform of $\exp^{-\alpha R}$, $R$ being the coordinate of the center of mass. That is
$$
\Theta(K)= N /(K^2+(\alpha \hbar c)^2)^2
\eqno(2.11)   
$$
      $K $ being measured in MeV's and $N$ being a normalization constant. As an example, consider $\alpha= 1 fm^{-1}$. In fig.1
      we plot the logarithm of the probabilities $p(n,l)$ for several $l$ values. As far as the wave-function is concerned
      few values of $n,l$ contribute to the expansion of eq.(2.7). An alternative way to illustrate the 
      pattern of convergence is illustrated in table 1, where we show the $\sum_{2n+l\leq N}p(n,l)(2l+1)$ as a function of $N$.
\renewcommand{\baselinestretch}{1}
\begin{figure}
\centering
\includegraphics[width=10.0cm,height=10.0cm,angle=0]{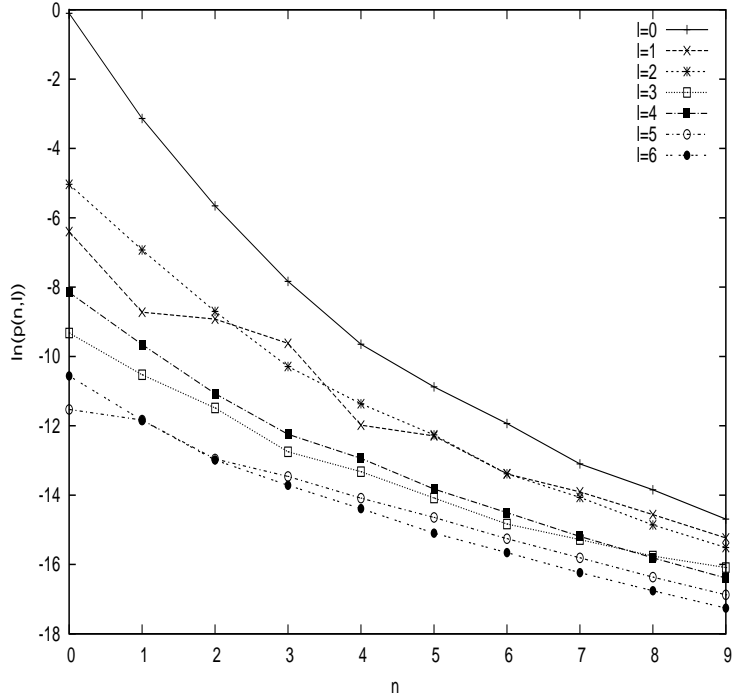}
\caption{Log of the probabilities of.eq.(2.10)as a function of the quantum number $n$ for several $l$ values.}
\end{figure}
\renewcommand{\baselinestretch}{2}
      A familiar nodal structure emerges by plotting the single-particle radial wave functions for several $l$-values.    
\renewcommand{\baselinestretch}{1}
\begin{table}
   \begin{tabular}{| c | c | c | c | }
          \hline
        N  & $ \sum_{2n+l \leq N} p_{n,l}(2l+1)$ & $ N $ & $ \sum_{2n+l \leq N} p_{n,l}(2l+1)$ \\
          \hline
        $   0   $ & $ 0.90275  $ & $  4   $ & $ 0.99584 $ \\
        $   1   $ & $ 0.90774  $ & $  5   $ & $ 0.99653 $ \\
        $   2   $ & $ 0.98374  $ & $  6   $ & $ 0.99868 $ \\   
        $   3   $ & $ 0.98485  $ & $  7   $ & $ 0.99905 $ \\   
          \hline
\end{tabular}
\caption { Accumulated probability as a function of N=max(2n+l).}
\end{table}
      In fig. 2 we show the first few radial wave-functions for $l=0$. The normalization implied by eq. (2.8) is
      $\int dk Q_{n,l}(k)Q_{n',l}(k)=\delta_{n,n'}$. In fig.3 and in fig. 4, we plot the radial   wave functions for $l=1$ and
      $l=6$ respectively. The nodal structure is clearly visible provided the label $n$ is associated to the 
      familiar harmonic oscillator radial quantum number.
\renewcommand{\baselinestretch}{1}
\begin{figure}
\centering
\includegraphics[width=10.0cm,height=10.0cm,angle=0]{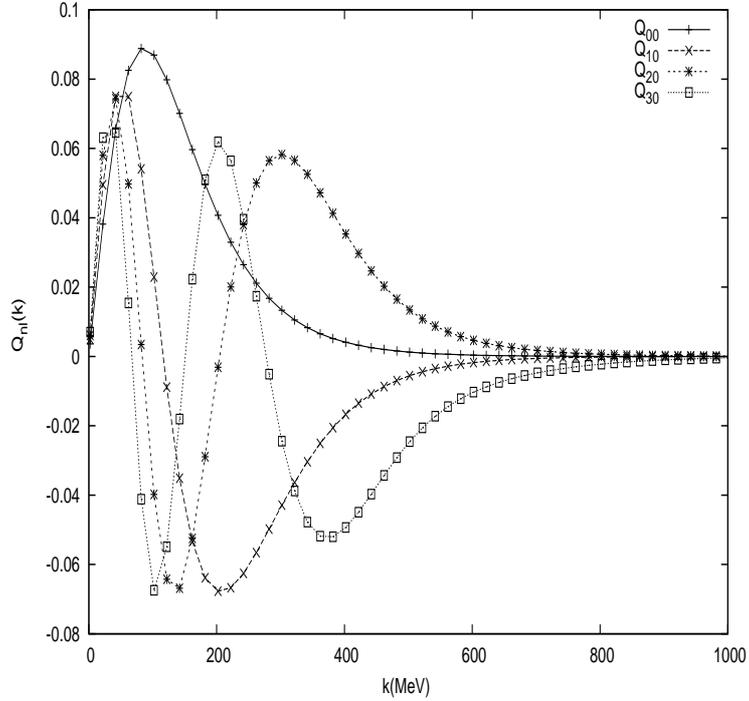}
\caption{ Radial wave functions $Q_{nl}(k)$ for $l=0$.}
\end{figure}
\renewcommand{\baselinestretch}{2}
\renewcommand{\baselinestretch}{1}
\begin{figure}
\centering
\includegraphics[width=10.0cm,height=10.0cm,angle=0]{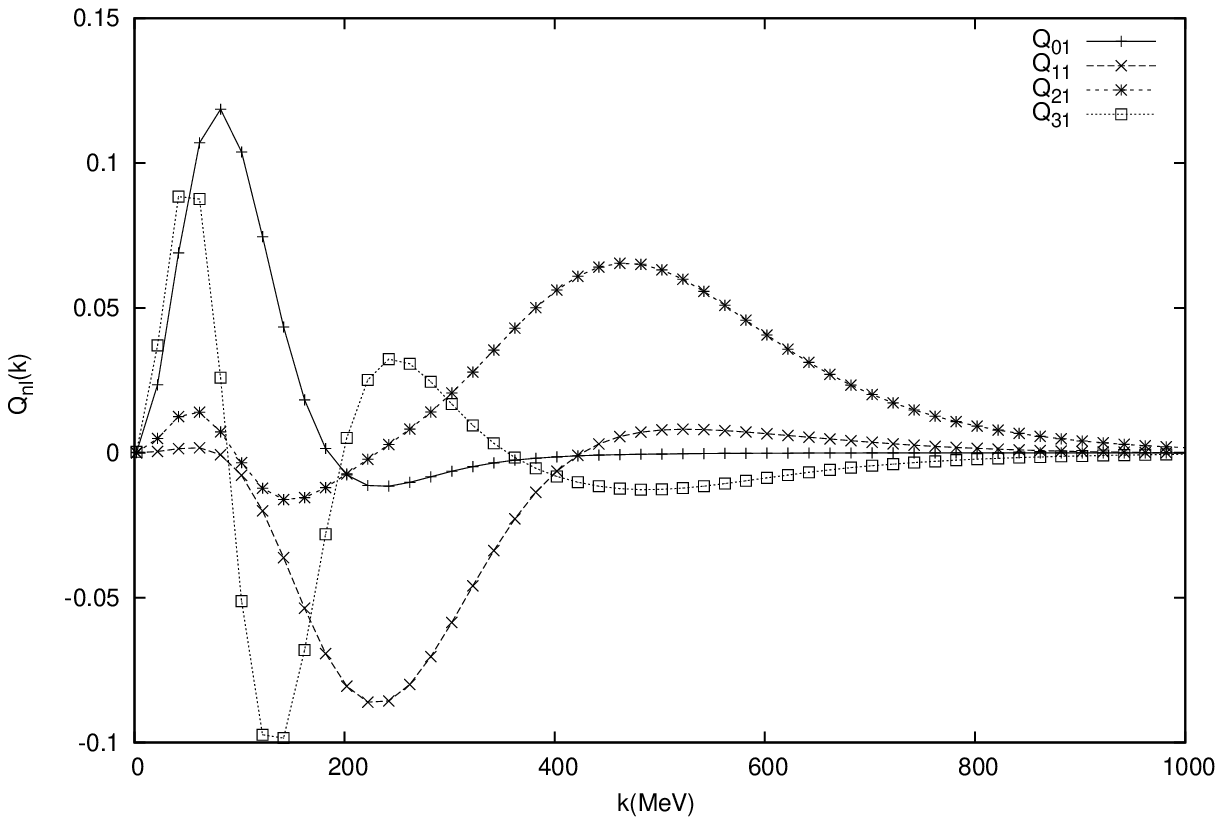}
\caption{ Same as fig. 2 for $l=1$.}
\end{figure}
\renewcommand{\baselinestretch}{2}
\renewcommand{\baselinestretch}{1}
\begin{figure}
\centering
\includegraphics[width=10.0cm,height=10.0cm,angle=0]{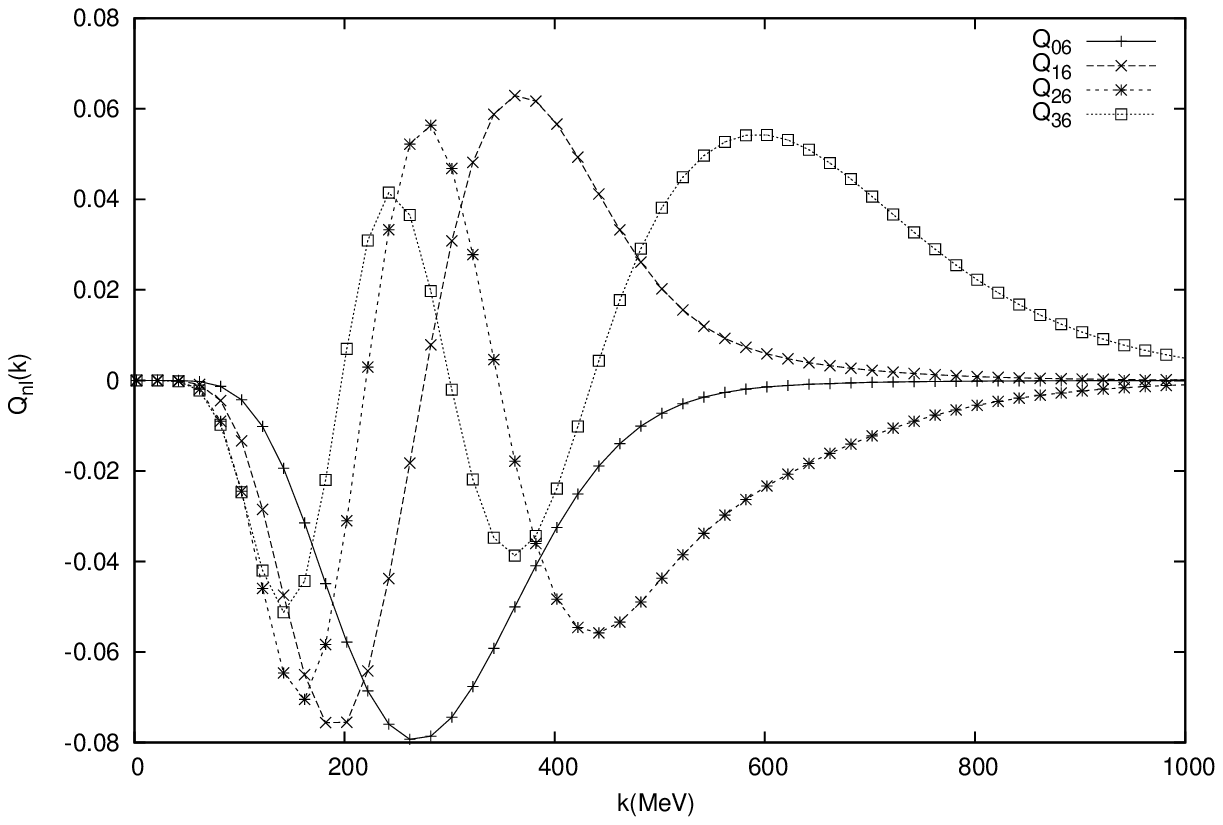}
\caption{ Same as fig. 2 for $l=6$.}
\end{figure}
\renewcommand{\baselinestretch}{2}
\par\noindent
 It should be stressed however that we do not have
      an oscillation theorem
      as for the h.o. radial wave-functions. The nodal structure of the radial wave-functions may depend on the original
      NN interaction. As a rule of the thumb, the "harder' the NN potential is at larger momentum transfer, the more
      distorted  the nodal structure can be.
\par
      These radial wave functions have the proper asymptotic behavior at large $r$ in coordinate space. One way to modify
      this asymptotic behavior in coordinate space is to use, instead of eq.(2.11), the Bessel-Fourier transform
      of a Gaussian for the center of mass.
\par   
\renewcommand{\baselinestretch}{1}
\begin{table}
   \begin{tabular}{| c | c | c | c | }
          \hline
        N  & $ \sum_{2n+l \leq N} p_{n,l}(2l+1)$ & $ N $ & $ \sum_{2n+l \leq N} p_{n,l}(2l+1)$ \\
          \hline
        $   0   $ & $ 0.71059  $ & $  4   $ & $ 0.96089 $ \\
        $   1   $ & $ 0.81441  $ & $  5   $ & $ 0.97420 $ \\
        $   2   $ & $ 0.90259  $ & $  6   $ & $ 0.98227 $ \\   
        $   3   $ & $ 0.93850  $ & $  7   $ & $ 0.98754 $ \\   
          \hline
\end{tabular}
\caption { Same as table 1 with $\alpha=2 fm^{-1}$.}
\end{table}
      We can modify the space extent of the radial wave-functions by modifying the parameter $\alpha$ in eq.(2.11).
      In table 2 we show the $\sum_{2n+l\leq N}p(n,l)(2l+1)$ as a function of $N$ for $\al=2 fm^{-1}$. 
      In this case the convergence
      to $1$ for increasing $N =max(2n+l)$ is slower. This is  not surprising since the Deuterium wave-function has a long tail
      in coordinate space and a "compressed" basis is less suited in an expansion of the Deuterium wave-function. In many-body
      calculations $\al$ is a variational parameter.
\renewcommand{\baselinestretch}{1}
\begin{figure}
\centering
\includegraphics[width=10.0cm,height=10.0cm,angle=0]{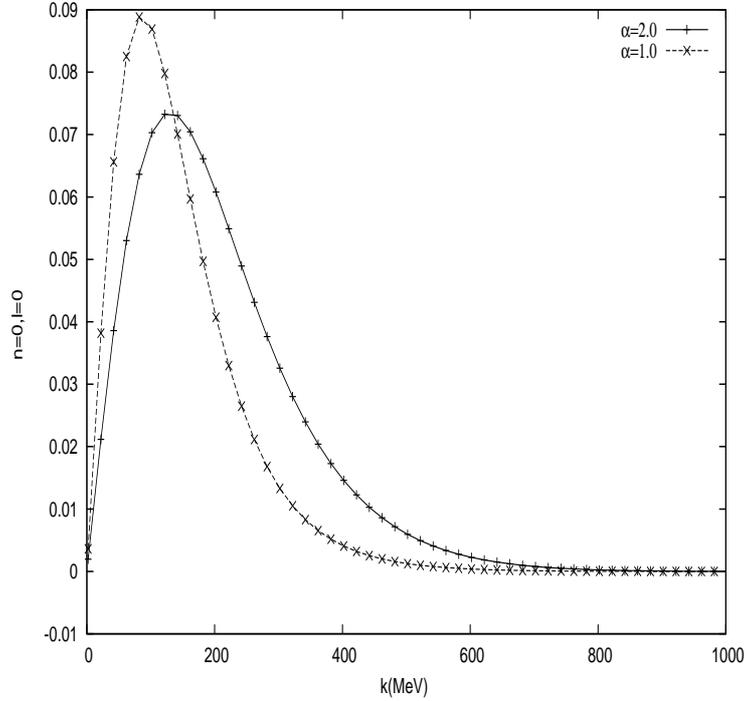}
\caption{ $Q_{0,0}$ for two different values of $\alpha$.}
\end{figure}
\renewcommand{\baselinestretch}{2}
\par
      In fig. 5 we show for comparison the radial $Q_{00}(k)$ evaluated at two different values of $\alpha$.
      The value of $\alpha$  has an analogous role of $\hbar\Om$ in the harmonic oscillator radial wave functions.
\par
      So far we have discussed a single-particle basis in the lab. frame derived from the "bare" NN interaction.
      One can ask whether these considerations are modified  if we soften the NN interaction with
      some renormalization procedure. We considered only the case of momentum cutoff to $\hbar c k_{max}=400 MeV$ in the
      frame of the center of  mass  ($V_{low k}$) for $\alpha=1 fm^{-1}$.
      The construction of the radial wave-functions in the lab. frame can be repeated
      as before. We found that the $Q_{nl}(k)$ show only minor differences especially for low values of $n,l$.
      Only at large values of $k\geq k_{max}$ and for large $n$ or $l$ in the lab. frame the $Q_{nl}$ show
      appreciable differences.
      Also the accumulated probabilities are very close to the ones of table 1. 
      Major differences might be found if the "bare" NN interaction is replaced by a much stronger one at large
      momentum transfer. In this work no renormalization steps have been taken. We use only the "bare" interaction.
\par
      The next step is the evaluation of the matrix elements of the interaction or  of the two-body
      matrix elements of the full Hamiltonian, in the new basis.
      This can be accomplished by first evaluating the matrix elements of the interaction (or of the
      the two-body matrix elements of the Hamiltonian) in a very large harmonic oscillators basis in the lab. 
      frame and then expanding  the matrix elements of the same operators in the new basis in terms of
      h.o. wave functions. See refs. [15] and [24] for a detailed discussion.
      By coupling the radial wave functions with the spin and angular part we can define the single particle
      basis as $|n,l,j,m>$. Let $|\overline n , \overline l,\overline j,\overline m>$ the h.o. counterpart and let us
      call ${P_{\overline n , \overline l}\over k}$ the corresponding h.o. radial wave functions. 
      If $a,b,c,d $ denote the set of quantum numbers $(n_a l_a j_a),(n_b l_b j_b),(n_c l_c j_c),(n_d l_d j_d)$
      in the new basis and $\overline a,\overline b,\overline c,\overline d $ the corresponding h.o. quantum numbers
      the transformation law for the angular momentum coupled two-body matrix elements of the interaction is
$$
<a,b J |V| c, d J>=\sum_{\overline a,\overline b,\overline c,\overline d }<a|\overline a><b|\overline b>
<\overline c|c> <\overline d|d> < \overline a\overline b J |V| \overline c \overline d J>
\eqno(2.12)
$$
       The overlaps $<a|\overline a>,...$ cannot change  the quantum numbers
      $l,j$ and they reduce to the radial integrals
$$
C^{(l)}_{n,\overline n}=\int_0^{\infty} dk Q_{n,l}(k)P_{\overline n,l}(k) 
\eqno(2.13)   
$$
      The degree of completeness of the selected h.o. space is assessed by the accumulated probabilities
$$
p^{(l)}_{n,\overline n}=\sum_{\overline\nu\leq\overline n} C^{(l)2}_{n,\overline\nu} 
\eqno(2.14)   
$$
      As an example let us consider $\hbar\om=14 MeV$ and $\alpha=1 fm^{-1}$. In table 3 we show the 
      amplitudes of eq.(2.12) as well as the accumulated probabilities for several values of $l$.
      The radial wave functions $Q_{n,l}/k$ were obtained from the "bare" NNLO-opt interaction      
      The expansion converges very fast for small $n,l$ and it is slower for large $n,l$. It is however faster than
      the expansion of the Coulomb-Sturm radial wave functions in terms of  a harmonic oscillator basis
      (compare table 3 with fig. 3 of ref.[15] for the Coulomb-Sturm case).
      Also, note that $max(2n+l)$ is the truncation parameter in many-body calculations.
      The convergence depends on the adopted values of $\alpha$ and $\hbar\om$. No optimization has been made in this example.
      In the actual calculations discussed in the next section an optimization has been performed.
      At this stage it is worth to notice that eq.(2.12) for a selected h.o.  subspace 
      works better for the interaction than for the full Hamiltonian. That is the kinetic energy terms
      might be poorly approximated in a "small" h.o. subspace. As done in refs. [12],[14], we use eq.(2.12)
      only for the interaction and evaluate all remaining terms of the Hamiltonian directly in the lab. frame.
\renewcommand{\baselinestretch}{1}
\begin{table}                 
   \begin{tabular}{| c |c |@{} r| c | c |@{}r | c | c |@{} r | c | c | r | c | c | r | c  }
          \hline
 $l,\overline n$&$ n $&$ C^{(l)}_{n,\overline n}$&$ p^{(l)}_{n,\overline n}$&$ n $&$ C^{(l)}_{n,
\overline n}$&$p^{(l)}_{n,\overline n}$&$
n $&$ C^{(l)}_{n,\overline n}$&$ p^{(l)}_{n,\overline n}$&$n$&$ C^{(l)}_{n,\overline n}$&$ p^{(l)}_{n,\overline n}$  \\
          \hline
$0,0$&$ 0 $&$ 0.983 $&$ 0.966 $&$ 1 $&$ 0.142 $&$ 0.020 $&$ 2 $&$ 0.110 $&$ 0.012$&$3 $&$ 0.034$&$0.001$\\
$0,1$&$ 0 $&$-0.116 $&$ 0.980 $&$ 1 $&$ 0.943 $&$ 0.909 $&$ 2 $&$-0.228 $&$ 0.064$&$3 $&$ 0.191$&$0.039$\\
$0,2$&$ 0 $&$ 0.132 $&$ 0.997 $&$ 1 $&$-0.138 $&$ 0.928 $&$ 2 $&$-0.867 $&$ 0.817$&$3 $&$-0.369$&$0.174$\\
$0,3$&$ 0 $&$-0.031 $&$ 0.998 $&$ 1 $&$ 0.243 $&$ 0.987 $&$ 2 $&$ 0.174 $&$ 0.847$&$3 $&$-0.729$&$0.706$\\
$0,4$&$ 0 $&$ 0.034 $&$ 0.999 $&$ 1 $&$-0.063 $&$ 0.991 $&$ 2 $&$-0.333 $&$ 0.958$&$3 $&$ 0.189$&$0.742$\\
$0,5$&$ 0 $&$-0.010 $&$ 0.999 $&$ 1 $&$ 0.082 $&$ 0.998 $&$ 2 $&$ 0.111 $&$ 0.970$&$3 $&$-0.397$&$0.899$\\
$0,6$&$ 0 $&$ 0.011 $&$ 1.000 $&$ 1 $&$-0.028 $&$ 0.998 $&$ 2 $&$-0.140 $&$ 0.990$&$3 $&$ 0.156$&$0.924$\\
$0,7$&$ 0 $&$-0.004 $&$ 1.000 $&$ 1 $&$ 0.032 $&$ 0.999 $&$ 2 $&$ 0.056 $&$ 0.993$&$3 $&$-0.213$&$0.969$\\
$0,8$&$ 0 $&$ 0.004 $&$ 1.000 $&$ 1 $&$-0.013 $&$ 0.999 $&$ 2 $&$-0.065 $&$ 0.997$&$3 $&$ 0.088$&$0.977$\\
$0,9$&$ 0 $&$-0.002 $&$ 1.000 $&$ 1 $&$ 0.014 $&$ 0.999 $&$ 2 $&$ 0.028 $&$ 0.998$&$3 $&$-0.115$&$0.991$\\
          \hline
$2,0$&$ 0 $&$ 0.976 $&$ 0.953 $&$ 1 $&$ 0.178 $&$ 0.032 $&$ 2 $&$ 0.114 $&$ 0.013$&$3 $&$-0.051 $&$ 0.003$\\
$2,1$&$ 0 $&$-0.129 $&$ 0.969 $&$ 1 $&$ 0.907 $&$ 0.854 $&$ 2 $&$-0.361 $&$ 0.143$&$3 $&$-0.135 $&$ 0.021$\\
$2,2$&$ 0 $&$ 0.159 $&$ 0.995 $&$ 1 $&$-0.215 $&$ 0.901 $&$ 2 $&$-0.762 $&$ 0.723$&$3 $&$ 0.523 $&$ 0.294$\\
$2,3$&$ 0 $&$-0.050 $&$ 0.997 $&$ 1 $&$ 0.266 $&$ 0.971 $&$ 2 $&$ 0.260 $&$ 0.791$&$3 $&$ 0.578 $&$ 0.628$\\
$2,4$&$ 0 $&$ 0.045 $&$ 0.999 $&$ 1 $&$-0.112 $&$ 0.984 $&$ 2 $&$-0.354 $&$ 0.916$&$3 $&$-0.167 $&$ 0.656$\\
$2,5$&$ 0 $&$-0.019 $&$ 1.000 $&$ 1 $&$ 0.101 $&$ 0.994 $&$ 2 $&$ 0.171 $&$ 0.945$&$3 $&$ 0.433 $&$ 0.843$\\
$2,6$&$ 0 $&$ 0.016 $&$ 1.000 $&$ 1 $&$-0.050 $&$ 0.997 $&$ 2 $&$-0.179 $&$ 0.977$&$3 $&$-0.146 $&$ 0.864$\\
$2,7$&$ 0 $&$-0.008 $&$ 1.000 $&$ 1 $&$ 0.045 $&$ 0.999 $&$ 2 $&$ 0.086 $&$ 0.985$&$3 $&$ 0.283 $&$ 0.945$\\
$2,8$&$ 0 $&$ 0.007 $&$ 1.000 $&$ 1 $&$-0.022 $&$ 0.999 $&$ 2 $&$-0.094 $&$ 0.994$&$3 $&$-0.083 $&$ 0.952$\\
          \hline
$4,0$&$ 0 $&$ 0.972 $&$ 0.946 $&$ 1 $&$-0.209 $&$ 0.044 $&$ 2 $&$ 0.088 $&$ 0.008$&$ 3 $&$-0.049 $&$ 0.002$\\
$4,1$&$ 0 $&$-0.152 $&$ 0.968 $&$ 1 $&$-0.873 $&$ 0.806 $&$ 2 $&$-0.438 $&$ 0.200$&$ 3 $&$-0.100 $&$ 0.012$\\
$4,2$&$ 0 $&$ 0.158 $&$ 0.993 $&$ 1 $&$ 0.276 $&$ 0.883 $&$ 2 $&$-0.683 $&$ 0.666$&$ 3 $&$ 0.527 $&$ 0.290$\\
$4,3$&$ 0 $&$-0.059 $&$ 0.997 $&$ 1 $&$-0.278 $&$ 0.960 $&$ 2 $&$ 0.284 $&$ 0.746$&$ 3 $&$ 0.551 $&$ 0.593$\\
$4,4$&$ 0 $&$ 0.046 $&$ 0.999 $&$ 1 $&$ 0.136 $&$ 0.978 $&$ 2 $&$-0.382 $&$ 0.892$&$ 3 $&$-0.035 $&$ 0.595$\\
$4,5$&$ 0 $&$-0.022 $&$ 1.000 $&$ 1 $&$-0.116 $&$ 0.992 $&$ 2 $&$ 0.182 $&$ 0.926$&$ 3 $&$ 0.488 $&$ 0.832$\\
$4,6$&$ 0 $&$ 0.018 $&$ 1.000 $&$ 1 $&$ 0.059 $&$ 0.995 $&$ 2 $&$-0.210 $&$ 0.970$&$ 3 $&$-0.050 $&$ 0.835$\\
$4,7$&$ 0 $&$-0.009 $&$ 1.000 $&$ 1 $&$-0.053 $&$ 0.998 $&$ 2 $&$ 0.096 $&$ 0.979$&$ 3 $&$ 0.324 $&$ 0.940$\\
          \hline
$6,0$&$ 0 $&$ 0.965 $&$ 0.932 $&$ 1 $&$-0.249 $&$ 0.062 $&$ 2 $&$-0.064 $&$ 0.004$&$ 3 $&$-0.038 $&$ 0.001$\\
$6,1$&$ 0 $&$-0.190 $&$ 0.968 $&$ 1 $&$-0.841 $&$ 0.769 $&$ 2 $&$ 0.463 $&$ 0.218$&$ 3 $&$-0.167 $&$ 0.029$\\
$6,2$&$ 0 $&$ 0.157 $&$ 0.992 $&$ 1 $&$ 0.312 $&$ 0.866 $&$ 2 $&$ 0.669 $&$ 0.665$&$ 3 $&$ 0.343 $&$ 0.147$\\
$6,3$&$ 0 $&$-0.065 $&$ 0.997 $&$ 1 $&$-0.297 $&$ 0.955 $&$ 2 $&$-0.211 $&$ 0.710$&$ 3 $&$ 0.614 $&$ 0.523$\\
$6,4$&$ 0 $&$ 0.049 $&$ 0.999 $&$ 1 $&$ 0.140 $&$ 0.974 $&$ 2 $&$ 0.433 $&$ 0.897$&$ 3 $&$ 0.107 $&$ 0.535$\\
$6,5$&$ 0 $&$-0.022 $&$ 0.999 $&$ 1 $&$-0.129 $&$ 0.991 $&$ 2 $&$-0.127 $&$ 0.913$&$ 3 $&$ 0.530 $&$ 0.816$\\
$6,6$&$ 0 $&$ 0.019 $&$ 1.000 $&$ 1 $&$ 0.060 $&$ 0.995 $&$ 2 $&$ 0.242 $&$ 0.971$&$ 3 $&$ 0.078 $&$ 0.822$\\
          \hline
\end{tabular}
\caption { Expansion coefficients of the $Q_{nl}$ in terms of the corresponding harmonic oscillator,
 $P_{\overline n l}$, see eqs.(2.12) and (2.13).}
\end{table}
\par  
     In the next section we shall optimize both values of $\hbar\om$ and $\alpha$ and shall study
     ${}^4He$ ${}^6He$ and ${}^6Li$. We shall show explicitly that the use
     of the basis described in this section leads to a decrease of the ground-state energies
     compared to the ones obtained using the harmonic oscillator basis with the optimal values
     of $\hbar\om$ except for ${}^4He$. We stress that we use only non-renormalized interactions. Hence $\hbar\om$ 
     is a variational parameter. The comparison will be made with calculations that have 
     the same computational burden.
\bigskip
\section{ Comparison between the h.o. and the new basis.}
\bigskip
{\it{ 3a. A brief recap of the HMD method.}}
\par
     Strictly speaking, the HMD method is a variational method based on the assumption
     that the nuclear wave function can be written as a linear combination of
     a number of Slater determinants (SD) with the option of projecting to good quantum numbers.
     These Slater determinants are of generic type and they
     are not orthogonal to each other, much in the same way of the Generator Coordinate Method.
     No assumption is made about the relevant degrees of freedom. The Slater determinants
     as well as the coefficients of the linear combination are determined only by variational 
     requirements. 
     The HMD method can take any input for the Hamiltonian which we schematically write as
$$
\HH = {1 \over 2} \sum _{i,j,k,l} H_{ijkl} \ad_i\ad_j a_la_k
\eqno(3.1)
$$
     The two-body matrix elements contain the "bare" two-body interaction, the intrinsic kinetic energy
     and the center of mass term $\beta(\HH_{cm}-3/2 \hbar\om)$, where $\beta$ is a coefficient
     and $\HH_{cm}$ is the harmonic oscillator Hamiltonian for the center of mass, 
     In eq.(3.1) $i,j,k,l$ are the single-particle quantum numbers $(n_i,l_i,j_i,m_i),...$ for 
     both neutrons and protons.
       We describe eigenstates
       as a linear superposition of Slater determinants of the most generic type
$$
| \psi>= \sum_{S=1}^{N_D} g_S \HP |U_S>
\eqno(3.2)
$$
       where $\HP$ is a projector to good quantum numbers (e.g. good angular momentum and parity)
       $N_D$ is the number of Slater determinants $|U_S>$ expressed as
$$
|U_S> = \oc_1(S)\oc_2(S)... \oc_A(S) |0>
\eqno(3.3)
$$
       the generalized creation operators $\oc_{\alpha}(S)$ for $\al=1,2,..,A$ are a linear combination
       of the creation operators $\ad_i$
$$
\oc_{\al}(S)=\sum_{i=1}^{N_s}U_{i,\al}(S)\ad_i  \;\;\;\;\;\al=1,...A
\eqno(3.4)
$$
       The complex coefficients $U_{i,\al}(S)$ represent the single-particle wave-function of the
       particle $\al=1,2,..,A$. We do not impose any symmetry on the Slater determinants (axial or other)
       since the $U_{i,\al}$  are variational parameters.
       These complex coefficients are obtained by minimizing the energy expectation values
$$
E[U]= { <\psi |\HH |\psi> \over <\psi |\psi>}
\eqno(3.5)
$$
       The coefficients $g_S$ are obtained by solving the generalized eigenvalue problem
$$
\sum_{S} <U_{S'} |\HP\HH | U_S> g_S = E \sum_{S} <U_{S'}|\HP| U_S> g_S
\eqno(3.6)
$$
       for the lowest eigenvalue $E$.
\par
       We consider a quasi-Newtonian minimization method. It is a generalization of
       the Broyden-Fletcher-Goldfarb-Shanno (BFGS) method (cf. for example ref.[26] and references in there).
       The variant we use is described in detail in ref. [27].
     The method starts with a small number of SD's (typically $1-5$) and determines the
     SD's  as well as the coefficients of the linear combination by minimizing the energy.
     Each SD is optimized individually. The process is repeated several times
     until all SD's have been optimized $N_T$ times. The number $N_T$ is such that
     an exit criterion is satisfied. The exit criterion is met when the energy changes
     less than a specified amount (typically $5 KeV$) between the $N_T-1$ and $N_T$ optimization.
     After the exit criterion is met, the number of SD's is increased by optimizing the last
     included SD. When the total number of SD's is large enough we repeat the optimization of
     all SD's one at a time. Typically, this way  we collect between $100$ and $200$ SD's.
     Typically, the full optimization is performed when the number of Slater determinants
     reaches the numbers $5,10,15,25,35,50,70,100,150,200,..$
     In eq.(3.1) if we select the harmonic oscillator basis, there are two possible inputs for the
     Hamiltonian matrix.  
     In the lab. frame the single-particle states satisfy
$$
a)\;\;\;\;\; 2 n+l \leq N_{2max}/2 
\eqno(3.7a)
$$
     where $N_{2max}$ is the largest total quantum number in the intrinsic frame.
    Or
$$
b)\;\;\;\;\; 2 n_1+l_1+ 2 n_2+l_2 \leq N_{2max}
\eqno(3.7b)
$$
    We stress that these are two possible truncations of the original "bare" two-body
    Hamiltonian. Eq.(3.7a) is referred as type (a) truncation and eq.(3.7b) as 
    type (b) of the original Hamiltonian. For larger and larger $ N_{2max}$ both approaches
    should hopefully converge to the same results.
    The type (a) may look a bit out of the ordinary and type (b) may seem preferable. 
    However we can argue as follows. Instead of harmonic oscillator single-particle
    wave functions we can take  other single-particle wave functions, for example
    the Coulomb-Sturm  wave functions or the ones considered in this work. 
    We cannot give to the radial quantum number
    the same meaning it has in the case of the  harmonic oscillator and actually condition
    eq.(3.7b) would seem a bit unjustified as there is no obvious reason why
    $2n+l$ for each particle, entering in the evaluation of the two-body matrix elements,
     should be related to each other. Type (a) truncation is more
    natural for single-particle basis other than harmonic oscillators.
    There is a further argument that one can offer. Consider an interaction either bare
    or softened with similarity renormalization group (SRG) methods, $V(q,q')$ with $q,q'$ being the relative momentum transfer
    between particles, and assume that we would like to evaluate directly in the lab. frame
    the two-body matrix elements of $V$ using the vector brackets; in such a case truncation
    (3.7b) would seem a bit unnatural and it would seem more reasonable to adopt the following
    criterion: all values of $(n,l)$ that contribute the most to the energies should be included.
    In this work, when using the h.o. representation, we select sometimes type(a) and sometimes type (b).
    When using the representation discussed in the previous section we use type (b) only in order to obtain
    the two-body matrix elements of the interaction (cf. eq. (2.12)). More explicitly,  using $N_{2max}=16\div22$ we first
     obtain the two-body matrix elements of the interaction of type (b)
     in the h.o. representation, then, using the expansion of the new basis in terms of h.o. single-particle
     radial wave functions, we obtain the matrix elements in the new basis with the restriction
     $2n+l\leq e_{max}$.  That is, using a basis other than the h.o.
     we use always the truncation of type type (a) where $N_{2max}/2$ is replaced by some maximum
     value $e_{max}$ of $2n+l$. In the new basis $2n+l$ does not have the meaning of major shell and it is simply
     a number that allows us to compare the results obtained using the h.o. basis of type (a) with the
     corresponding ones obtained in the new basis. Differently stated, we compare truncation of type (a)  
     for the h.o. representation with the analogous truncation in the new basis. 
     Note that the truncation of the Hilbert space used in this work is very different from the usual one
     used in shell model calculations (refs. [15],[16]). The $N_{max}$ truncation used in shell model calculations
     refers to many-body configurations, i.e. total maximum number of oscillator quanta minus the minimal one.
\par
\bigskip
\bigskip
{\it{ 3b. Some numerical results.}}
\par
     In all cases treated in this subsection we consider up to $l=5$ for the single-particle orbital
      angular momentum.
     The Slater determinants were determined by minimizing the energies using a projector to good
      z-projection of the angular momentum and parity $J_z^{\pi}$  as explained in previous subsection.
\par
      All energies depend on $e_{max}=max(2n+l)$ and the total number of employed Slater determinants $N_D$.
      We optimize $\hbar\om$ for the calculations using the h.o. representation and $\alpha$ in the
      new LDB representation. The optimization is performed with few Slater determinants and mostly the value
      of $\hbar\om$ or $ \alpha$ is kept for the rest of the calculations.
      The results are shown in table 4.
\begin{table}
	\begin{tabular}{| c | c | c | c | c | c | c | c |}
                \hline
                $ {}^AZ  $ & $ e_{max} $ & $ N_D $ & $ \hbar\om $ & $ E_{ho}(MeV) $ & $\al(fm^{-1})$ & $  E_{LDB}(MeV) $ &
                $ E^{CM}_{LDB}(MeV) $  \\ \hline
  $ {}^4He $ & $ 6 $ & $ 150 $ & $28 $ & $ -27.444  $ & $ 4.0 $ & $ -27.145 $ & $ 0.039$ \\
  $ {}^4He $ & $ 7 $ & $ 150 $ & $28 $ & $ -27.488  $ & $ 4.0 $ & $ -27.386 $ & $ 0.029$ \\
\hline
  $ {}^6Li $ & $ 5 $ & $ 150 $ & $20 $ & $ -25.649  $ & $ 2.55$ & $ -25.737 $ & $ 0.215$ \\
  $ {}^6Li $ & $ 6 $ & $ 150 $ & $20 $ & $ -26.648  $ & $ 2.75$ & $ -26.847 $ & $ 0.147$ \\
  $ {}^6Li $ & $ 7 $ & $ 150 $ & $20 $ & $ -27.229  $ & $ 2.75$ & $ -27.665 $ & $ 0.123$ \\
\hline
  $ {}^6He $ & $ 5 $ & $ 150 $ & $16 $ & $ -21.506  $ & $ 2.25$ & $ -22.822 $ & $ 0.187$ \\
  $ {}^6He $ & $ 6 $ & $ 150 $ & $16 $ & $ -22.919  $ & $ 2.25$ & $ -23.786 $ & $ 0.117$ \\
  $ {}^6He $ & $ 7 $ & $ 150 $ & $16 $ & $ -23.885  $ & $ 2.25$ & $ -24.357 $ & $ 0.112$ \\
                \hline
        \end{tabular}
 \caption { Comparison between results obtained using the h.o. basis and the LDB.
 Note that $e_{max}$ and $  N_D$ are not sufficient to obtain convergence.
 The last column gives the expectation values of $\bet(H_{cm}-3/2 \hbar\om)$
 for $\bet=0.5$ using LDB.}
\end{table}
\par
      Note that the use of a basis different from the h.o. does not introduce strong center of mass
      excitations. Actually, the residual $<\bet(H_{cm}-3/2 \hbar\om)>$ decreases with larger
      $e_{max}$. Also, most of the times, we did not optimize the value of $\hbar\om$ in the
      center of mass Hamiltonian. We have used most of the times the value of $\hbar\om$  
      optimized in the h.o. representation. Apart from the case of ${}^4He$ the results are
      encouraging. The computational cost of the two representation is roughly the same 
      and the LDB does not require additional many-body calculations as in the natural orbits
      approach. It is natural to ask whether the decrease in the energies remains as we consider
      heavier nuclei. We made a test with ${}^{40}Ca$ for $e_{max}=5$ ($6$ major shells), up to
      $35$ Slater determinants using both the h.o. representation and the LDB introduced in this work.
\begin{figure}
\centering
\includegraphics[width=10.0cm,height=10.0cm,angle=0]{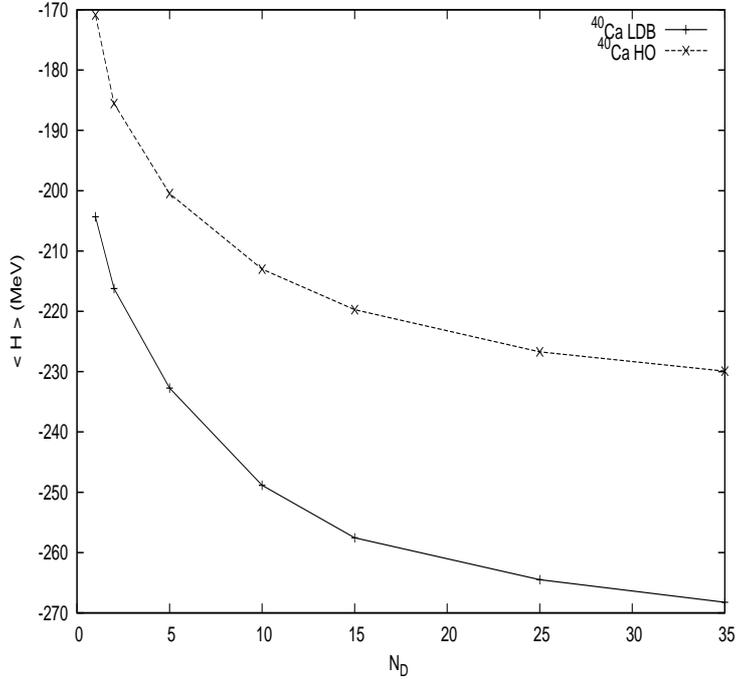}
\caption{Energies for ${}^{40}Ca$  using 6 major shell ($e_{max}=5$) up to $N_D=35$ Slater determinants in the LDB and H.O.
representations,}
\end{figure}
\renewcommand{\baselinestretch}{2}

      In fig. 6 we compare the energies as a function of $N_D$ obtained using the h.o. and the LDB.
      The result is quite encouraging. With $35$ Slater determinants, LDB lowers the energies
      by about $38 MeV$, compared to the standard h.o. representation. We stress that convergence is not
      reached and a much larger number of $N_D$ is needed as well as a larger number of major shells.
      As such, fig. 6 should be regarded as highly preliminary.
      Note, also, that  $\hbar\om$ has been optimized to the h.o. representation and kept
      the same in the center of mass Hamiltonian in the LDB.
\bigskip
\section {Conclusions.}
      In this work we have presented a new single-particle basis for many-body calculations.
      This basis is extracted from a two-body problem, by adding a localizing wave function
      in the center of mass coordinate to the intrinsic two-body eigenstate. The full wave function
      is analyzed in terms of Legendre polynomials and rewritten as sum of products of single-particle
      wave functions. Essentially this basis is constructed by diagonalization of  a two-body wave function
      rather than a single-particle Hamiltonian. For ${}^6Li$ and ${}^6He$ it gives lower values
      of the energy when compared with the values obtained using the harmonic oscillator representation.
      Moreover, in a preliminary study, it seems to be ideal for medium mass systems.

\vfill
\eject
\end{document}